\documentclass[aps,pre,floatfix,twocolumn,showpacs,showkeys,amsmath,amssymb,superscriptaddress,10pt]{revtex4-1}

\usepackage{amsthm}
\usepackage{bm}
\usepackage{epsfig}
\usepackage{verbatim}
\usepackage{graphicx}

\newcommand{\beq}{\begin{equation}}
\newcommand{\eeq}{\end{equation}}
\newcommand{\beqa}{\begin{eqnarray}}
\newcommand{\eeqa}{\end{eqnarray}}

\usepackage{xcolor}

\begin{document}

\title{A three-state model with competing antiferromagnetic and pairing interactions} 

\author{Sergio A. Cannas}
%\email{cannas@famaf.unc.edu.ar}
\affiliation{Facultad de
Matem\'atica, Astronom\'{\i}a, F\'{\i}sica y Computaci\'on, Universidad Nacional
de C\'ordoba, Instituto de F\'{\i}sica Enrique Gaviola (IFEG-CONICET)\\ Ciudad Universitaria, 5000 C\'ordoba, Argentina}
\author{Daniel A. Stariolo}
%\email{stariolo@if.uff.br}
\affiliation{Departamento de F\'{\i}sica,
Universidade Federal Fluminense and
National Institute of Science and Technology for Complex Systems\\
 Campus Praia Vermelha, 24210-346 Niter\'oi, RJ, Brazil}

\date{\today}

\begin{abstract}
  Motivated by the rich phase diagram of the high temperature superconductors, we introduce a toy model with
  three state variables which can be interpreted as two state particles and holes. The Hamiltonian has
  a term which favors antiferromagnetism and an additional competing interaction which favors  bonding between pairs of antiparallel
  spins mediated by holes. For low concentration of holes the dominant interaction between particles has antiferromagnetic character, leading to an
  antiferromagnetic phase in the temperature-hole concentration phase diagram, qualitatively similar
  to the antiferromagnetic phase of doped Mott insulators. For growing concentration
  of holes antiferromagnetic order is weakend and a phase with a different kind of order mediated by holes appears. This last phase has the form of a dome in the T-hole concentration plane. The whole phase diagram
  resembles those of some families of high $T_c$ superconductors. We compute the phase diagram in the
  mean field approximation and characterize the different phase transitions through Monte Carlo
  simulations.
\end{abstract}

%\pacs{05.50.+q,64.60.Cn,75.40.-s}
%\keywords{}

\maketitle

\section{Introduction}
One of the frontiers of modern condensed matter physics is the description of phase transitions in
complex many body systems that challenge the well known theories of classical fluids or the well
established quantum
theories of Fermi or Bose liquids.
In particular, when two or more competing interactions are present, usually strong correlations play a
significant role in the physics at low temperatures and the phase diagram can show a rich variety
of different phases, with different symmetries and thermodynamic properties. Competing interactions are
common in nature, an important example are magnetic systems in which different types of magnetic interactions
usually coexist like exchange, dipolar, Dzyaloshinsky-Moriya and Zeeman~\cite{PhysRevLett.117.157205,
PhysRevB.93.184413,PhysRevB.81.094407,RevModPhys.85.1473,Saratz2016}. Each of these terms usually lead to
a particular kind of order at low temperatures, but when at least two of them are simultaneously present a much
richer behaviour appears. Competition between interactions can lead to pattern formation and phases with
complex symmetries~\cite{PhysRevLett.104.077203,PhysRevB.75.104431,PhysRevB.84.014404,BaSt2009}.
High temperature superconductors appear as another example
of extrememly complex and fascinating systems. Since their discovery in the 80's thousands of papers have been
published on the subject, but after 30 years of very intense research both the experimental and theoretical
situations are still not satisfactory~\cite{RevModPhys.84.1383,Dai2012,RevModPhys.87.457,Keimer2015,Vojta2009}. From the theoretical perspective, perhaps the biggest open
question is the nature of the microscopic pairing mechanism which leads electrons to form Cooper pairs in the
high $T_c$ superconductor families of compounds~\cite{RevModPhys.84.1383,Boke1501329}. It is accepted that the BCS theory, so successful at explaining
superconductivity in the ``usual'' superconductors~\cite{Tinkham,Annett}, does not explain the superconductiong behavior of the cuprates or the iron based families.
In the cuprates or iron based superconductors strong electronic correlations seem to be at work forcing the
theoretical description of the relevant physics to go beyond the classic BCS theory. One of the striking effects
of the strong electronic correlations is the simultaneous relevance and consequent competition between
different degrees of freedom, leading to possibly different kind of orders associated to electron density
, spin and orbital degrees of freedom. It is not even completely clear when these different orders compete with each other
or instead in some sense ``catalyse'' the appearence of superconductivity. Sometimes this is refereed to as
``intertwinned orders''~\cite{BeFrKiTr2009,RevModPhys.87.457}. From this simple perspective it is immediately clear that the phase diagrams of these
systems are extremely rich and complex. In fact, both experimental and theoretical work usually focus on one
or few relevant observables in limited regions of parameter space. An overall understanding of the phase
diagram, collecting all or most of the pieces at work is still not available.

The more well known families of high
$T_c$ superconductors are the ``cuprates''~\cite{RevModPhys.87.457,Keimer2015} and the iron based ones or ``pnictides''~\cite{Dai2012,FernandesChubukov2017}.
Both families show similar thermal properties, although not identical. In both cases superconductivity arises
by doping with electrons or holes the so called parent compounds. The undoped parent compounds are insulators
at low temperatures, showing antiferromagnetic order at half filling. As long as the doping is introduced in
the system antiferromagnetism gradually gets weaker and eventually dissapears. Beyond some degree of doping,
which depends on the compound, a superconducting phase appears at very low temperatures. By increasing the
doping the superconducting phase extends to higher temperatures until a maximum $T_c$ is obtained at an ``optimal''
doping level. Beyond optimal doping the critial superconducting temperature goes down and eventually goes to
zero at another critical doping level. This is the well known ``superconducting dome''. In general, in the cuprate
superconductors antiferromagnetism dissapears for a doping level less than that at which superconductivity appears,
the antiferromagnetic phase and the superconducting dome do not intersect in the temperature-doping plane. On
the other side, in the iron based superconductors the antiferromagnetic and superconducting phases usually appear
superposed, leading to a region of coexistence between antiferromagnetism and superconductivity. In the same
temperature-doping plane of the phase diagram other phases, associated with other degrees of freedom and
different symmetries usually appear, like modulated electronic and spin orders, called stripes and nematic
phases~\cite{KivFrad2003,Vojta2009,Fradkin}. Structural transitions in the crystal symmetry of the compounds are also observed probably
being relevant
for the emergence of superconductivity or other types of order seen in the phase diagram.
The common presence of quenched disorder in the samples also leads to freezing of degrees of freedom
and spin glass like behavior in some cases. From this crude exposition of the thermal phenomenology of high $T_c$
superconductors one can immediately conclude that obtaining a complete phase diagram is a formidable task.

From a statistical mechanics point of view it is usually very helpful to develop a simplified model which
captures some of the relevant properties of a real system. This approach has proved to be extremely
successful in the context of critical phenomena after the appearance of the concept of universality.
In a nutshell, universality means that the behavior of the order parameters and the associated responses
to the conjugate fields of any system near a
continuous phase transition depend only on a few relevant charateristics of the system, basically the
symmetries of the order parameter and the dimensionality of space. In particular, microscopic details of
the interactions are not relevant regarding the critical behaviour. A well known example is the Blume-Emery-Griffiths (BEG)
model~\cite{BEG} which, despite its extreme simplicity, completely describes the phase diagram global topology
of $^3$He and $^4$He mixtures, including the right order of the phase transitions. Extensions of the BEG model has also been
proposed to describe some aspects of superconductivity~\cite{PhysRevB.78.184519}. 
With this in mind and motivated by the
rich complexity of the phase diagram of high $T_c$ superconductors,
in this paper we introduce a very simplified model which captures the topology of the antiferromagnetic and
superconductor phases of those systems. To our knowledge, these two phases have only been obtained from
a single model Hamiltonian in very few cases e.g. \cite{PhysRevLett.104.057006,LI2016925} and
references therein (see also \cite{PhysRevB.78.184519}), and because of the complexity of the models involved and
the different nature of the order parameters a
detailed analysis of the phase transitions have not been done yet. Then, although  the present
model is clearly too simplified to describe the complex physics of the cuprates or iron based superconductors, we nevertheless
think it can be useful to think on the universal or robust thermal properties which can lead to the particular
phase diagrams of such systems.

\section{The model}

The model is defined by a set of three-state classical variables $S_{i,j}=0,\pm 1$, $i,j=1\ldots L$, interacting in a square lattice of $N=L\times L$ sites, where $S_{i,j}=0$ represents a hole and the two other states $S_{i,j}=\pm 1$ may be thought of as
representing the spin states of an electron.
%Hence the occupation variable is $n_{i,j}=S_{i,j}^2=0,1$.
The Hamiltonian has two terms which compete with each other: a nearest-neighbour interaction of strength
$J_A$ which favors antiferromagnetic order, and a three-site interaction mediated by a hole, of strength
$J_B$. If the concentration of holes is zero this interaction is not present and the system shows only
an antiferromagnetic phase of N\'eel type at low temperatures. The presence of a hole favors another kind of
antiferromagnetic order, this time between third neighbours along the two principal directions of the square
lattice with a hole in between. The Hamiltonian is given by

\begin{widetext}
\begin{equation} \label{hamilton}
{\cal H}=   \sum_{i=1}^L \sum_{j=1}^L \left[\frac{J_A}{2} S_{i,j} \left(S_{i+1,j} +S_{i-1,j}+S_{i,j+1}+S_{i,j-1}\right)+ J_B (1-S_{i,j}^2) \left(S_{i+1,j} S_{i-1,j}+S_{i,j+1} S_{i,j-1}\right)\right]
\end{equation}
\end{widetext}

\noindent where $J_A>0$, $J_B>0$. The motivation behind the second term is to induce a kind of ``pairing''
between couples of particles with up and down spins mediated by a hole.

\begin{figure}[ht!]
%\centering
\includegraphics[scale=0.4]{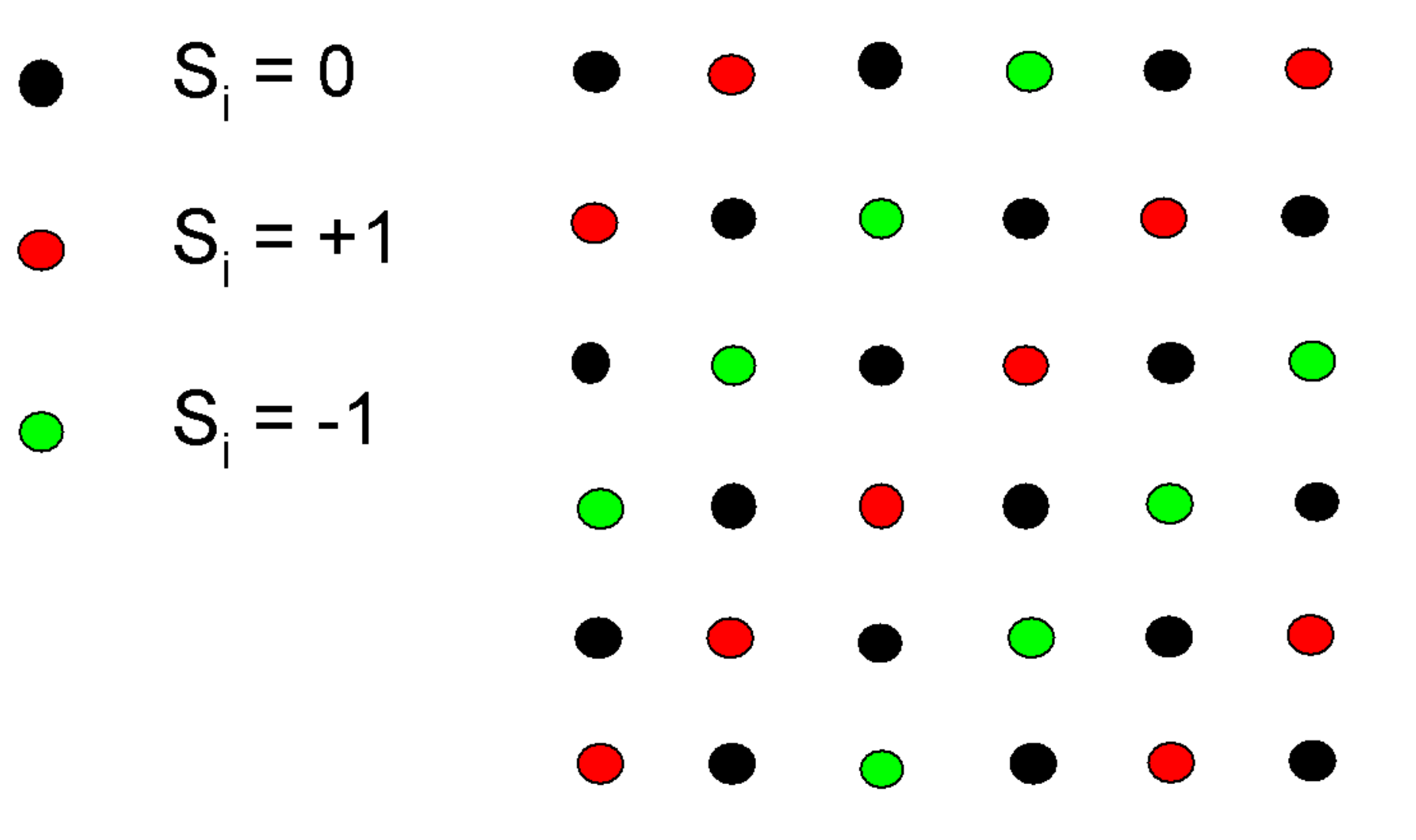}
\caption{(Color online) Super-Antiferromagnetic (SAF) pattern.}
\label{fig:saf}
\end{figure}

%and periodic boundary conditions are assumed.
At zero temperature there are two possible ground states: the usual antiferromagnetic N\'eel state and a
``super-antiferromagnetic'' (SAF) state, as shown in Figure \ref{fig:saf}.
Then, to study the phase diagram and phase transitions present in the system
we considered two order parameters: the usual antiferromagnetic order parameter (staggered magnetization)
and an orientational parameter which characterizes the super-antiferromagnetic order defined
as:
\begin{widetext}
\begin{equation}
{\cal O}= \frac{1}{2N}\sum_{i=1}^L \sum_{j=1}^L   S_{i,j}\left[S_{i+1,j+1}+S_{i-1,j-1}-S_{i-1,j+1}-S_{i+1,j-1}  \right]
\label{Ohv}
\end{equation}
\end{widetext}
This order parameter is able to capture emergent order induced by the second interaction term in
\eqref{hamilton}.

\section{Mean field phase diagram}

The grand partition free energy density is given by:
\begin{equation}
f=-\frac{1}{\beta N} \ln \left[ {\rm Tr}\, e^{-\beta H}\right]
\end{equation}
\noindent where $H={\cal H}-\mu M$ and
\begin{equation}
M=\sum_{i=1}^L \sum_{j=1}^L S_{i,j}^2
\end{equation}
\noindent is the total number of particles and $\mu$ the corresponding chemical potential.

A mean field free energy $\Phi$ can be obtained using the Bogoliubov inequality:
\begin{equation}
f\leq \Phi = f_0 + \frac{1}{N}\langle H-H_0 \rangle_0
\end{equation}
\noindent where $H_0$ is a reference non-interacting variational Hamiltonian with
\begin{equation}
f_0=-\frac{1}{\beta N} \ln Z_0
\end{equation}
\noindent and
\begin{equation}
Z_0={\rm Tr}\, e^{-\beta H_0}
\end{equation}
its partition function. $H_0$ assumes the general form:
\begin{equation}
 H_0=-\sum_{i=1}^L \sum_{j=1}^L \eta_{i,j}\, S_{i,j} - \sum_{i=1}^L \sum_{j=1}^L \mu_{i,j} S_{i,j}^2,
\end{equation}
\noindent where $\mu_{i,j}$ and $\eta_{i,j}$ are local variational parameters. Then, $Z_0=\prod_{i,j} Z_0^{i,j}$ with:
\begin{equation}
Z_0^{i,j}=\sum_{S=0,\pm 1} e^{\beta\eta_{i,j}S +\beta\mu_{i,j} S^2}= 1+2\, e^{\beta\mu_{i,j} }\, \cosh \left(\beta\eta_{i,j} \right),
\end{equation}

and

\begin{eqnarray}
  \langle S_{i,j} \rangle_0&=&\frac{1}{Z_0^{i,j}}\sum_{S=0,\pm 1} \,S e^{\beta\eta_{i,j}S +\beta\mu_{i,j}  S^2}\nonumber \\
  &=& \frac{2}{Z_0^{i,j}}\, e^{\beta\mu_{i,j} }\, \sinh \left(\beta\eta_{i,j} \right)
\end{eqnarray}

\begin{eqnarray}
  \langle S^2_{i,j} \rangle_0&=&\frac{1}{Z_0^{i,j}}\sum_{S=0,\pm 1} \,S^2 e^{\beta\eta_{i,j}S +\beta\mu_{i,j}  S^2}\nonumber \\
  &=& \frac{2}{Z_0^{i,j}}\, e^{\beta\mu_{i,j} }\, \cosh \left(\beta\eta_{i,j} \right)
\end{eqnarray}

\subsection{Antiferromagnetic solution}

We consider first the antiferromagnetic solution $\eta_{i,j}=\pm \eta$, according to which sublattice the site
of coordinates ${i,j}$ belongs, and $\mu_{i,j} =\mu$.
This implies a constant density:

\begin{eqnarray}
  \rho=\langle S^2_{i,j} \rangle_0&=&\frac{2\,e^{\beta\mu}\, \cosh \left(\beta\eta \right)}{1+2e^{\beta\mu}\, \cosh \left(\beta\eta \right)}\nonumber \\
  &=&  \frac{2\, \cosh \left(\beta\eta \right)}{e^{-\beta\mu}+2\, \cosh \left(\beta\eta \right)}
  \label{eq:rhoAF}
\end{eqnarray}
\noindent and a staggered manetization $\langle S_{i,j} \rangle_0=\pm m$ with
\begin{equation}
  m=\frac{2\,e^{\beta\mu}\, \sinh \left(\beta\eta \right)}{1+2\,e^{\beta\mu}\, \cosh \left(\beta\eta \right)}
  =\frac{2\, \sinh \left(\beta\eta \right)}{e^{-\beta\mu}+2\, \cosh \left(\beta\eta \right)}.
\label{eq:mAF}
\end{equation}
It is easy to obtain $Z_0^{i,j}=\frac{1}{1-\rho}$ and therefore $f_0=\frac{1}{\beta} \ln (1-\rho)$.

\noindent Then, the variational free energy density of the antiferromagnetic phase reads:

\begin{widetext}
\begin{equation}
\phi_{AF}=\Phi_{AF}/N=\frac{1}{\beta} \ln{(1-\rho)} + m^2 \left(-2J_A+(1-\rho)J_B \right) + m \eta.
\end{equation}
\end{widetext}

\begin{figure}[ht!]
%\centering
\includegraphics[scale=0.7]{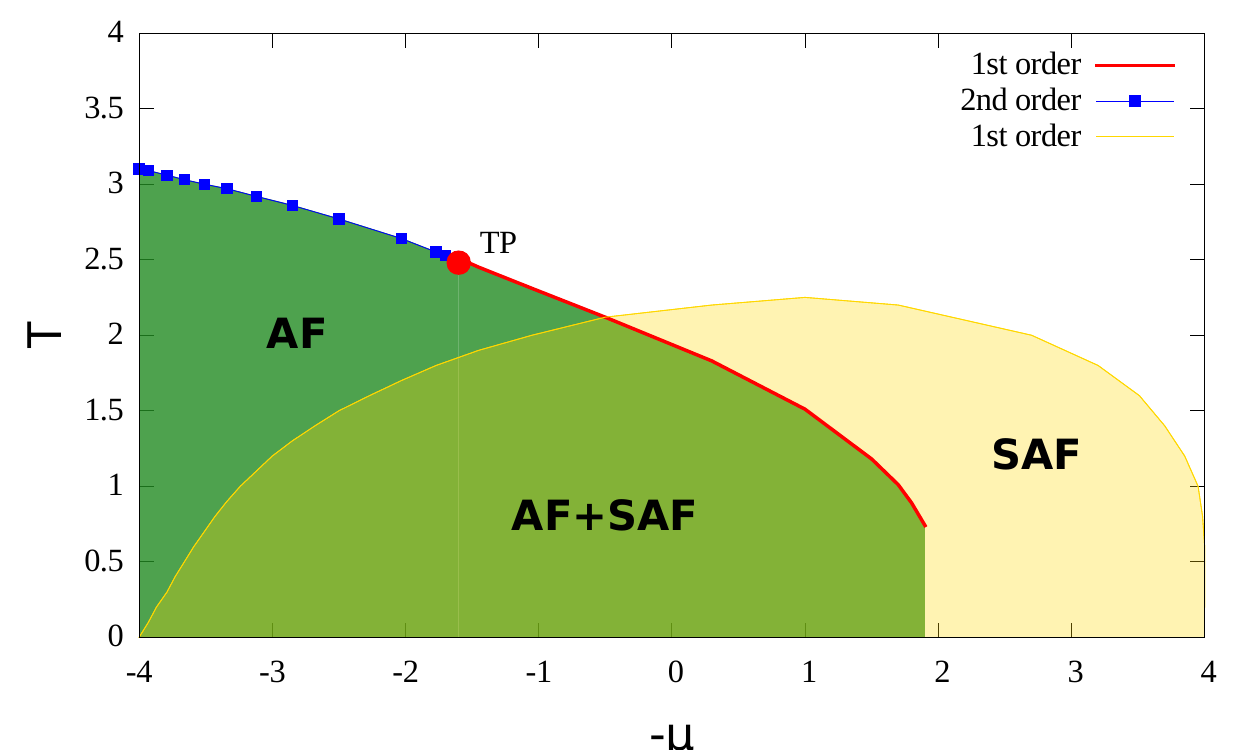}
\caption{(Color online) Mean field phase diagram for $J_B=2J_A$.}
\label{fig:meanfield}
\end{figure}

Solving for the stationary condition on the variational parameter $\eta$,
$\frac{\partial \phi_{AF}}{\partial \eta}=0$ gives:
\begin{equation}
  \eta = m \left\{ 4J_A + (1-\rho)J_B \left[\frac{m^2}{\rho-m^2}-2\right] \right\}.
  \label{eq:eta}
\end{equation}
Inserting this result in (\ref{eq:rhoAF}) and (\ref{eq:mAF}) we obtain two coupled equations for the order parameters $\rho$
and $m$ which must be solved self-consistently. From (\ref{eq:eta}) $m=0$ is always a solution.
In the limit $\mu\to\infty$ we have $\rho=1$ and $m = {\rm tanh}(4J_A \beta m)$, as expected. The
numerical solution of the self consistent equations for $m$ and $\rho$ suggests that for large positive
values of the chemical potential $\mu$ there is a line of continuous transitions between the
antiferromagnetic and paramagnetic phases, which ends at a tricritical point. For locating the
tricritical point we did a Landau expansion of the free energy density:
\begin{equation}
  \phi_{AF} = \phi_0 +A\,m^2 + B\,m^4 + C\,m^6 + \ldots
\end{equation}
The second order transition line is defined when $A(\mu,T)=0$. This gives:
    \begin{equation}
      \frac{1}{\rho}=\beta \left[ 4J_A-2J_B(1-\rho)\right]
    \end{equation}
    Changing variables to $T=1/\beta$ (in units of the Boltzmann constant $k_B$) and defining the density of holes $x=1-\rho$,
    we obtain the critical line $T(x)$:
    \begin{equation}
      T(x) = (1-x)\left[ 4J_A-2J_Bx\right].
      \label{eq:AFphasediag}
    \end{equation}
 This should
    be the phase diagram of the antiferromagnetic phase assuming a continuous transition for any
    $x$. Nevertheless, it is possible to show that the continous transition line ends at a
    tricritical point $(\mu^t,T^t)$. This is identified imposing the simultaneous vanishing
    of the prefactors $A(\mu,T)=0$ and $B(\mu,T)=0$. For the case $J_B=2J_A$
    the tricritical point can be found numerically to be at $(\mu^t,T^t)=(1.77,2.55)$.
    For $\mu < \mu^t$ the line of transitions is of first order. This is illustrated in the
    $T-\mu$ phase diagram in Figure \ref{fig:meanfield}.

\subsection{Super-Antiferromagnetic solution}

The SAF ground state has the symmetry shown in Figure \ref{fig:saf}. The system is divided in two
interpenetrated sublattices A and B (black and red-green sites in the figure). Sublattice A
has a uniform low-density, low-magnetized state. Sublattice B has a striped state (red and green
sites in Figure \ref{fig:saf}). Then, for the SAF phase we choose the variational parameters to be
$\mu_{i,j}=\mu+\delta$ and $\eta_{i,j}=0$ for all sites $(i,j)$ in the sublattice A. For all
sites $(i,j)$ in the sublattice B we set   $\mu_{i,j}=\mu$ and $\eta_{i,j}=\pm \eta$ for sites
belonging to alternated columns. Then, for sites belonging to the sublattice A
$\langle S_{i,j} \rangle_0^A  =  m_A=0$ and:
\begin{equation}\label{eq:rhoA}
  \langle S^2_{i,j} \rangle_0^A  =  \rho_A(\delta)=\frac{2 e^{\beta(\mu+\delta)}}{1+2 e^{\beta(\mu+\delta)}}
  =\frac{2}{e^{-\beta(\mu+\delta)}+2}.
\end{equation}
\noindent  For sites belonging to the sublattice B:
\begin{eqnarray}
  \langle S_{i,j} \rangle_0^B  =  \pm m_B(\eta)&=&\pm
  \frac{2 e^{\beta\mu}\, \sinh \left(\beta\eta \right)}{1+2 e^{\beta\mu}\, \cosh \left(\beta\eta \right)}\nonumber \\
  &=&\pm \frac{2\, \sinh \left(\beta\eta \right)}{e^{-\beta\mu}+2\, \cosh \left(\beta\eta \right)}\label{eq:mB},
\end{eqnarray}
and
\begin{eqnarray}
  \langle S^2_{i,j} \rangle_0^B=\rho_B(\eta)&=&\frac{2 e^{\beta\mu}\, \cosh \left(\beta\eta \right)}
          {1+2 e^{\beta\mu}\, \cosh \left(\beta\eta \right)} \nonumber \\
          &=&\frac{2 \, \cosh \left(\beta\eta \right)}
          {e^{-\beta\mu}+2 \, \cosh \left(\beta\eta \right)}\label{eq:rhoB}.
\end{eqnarray}

In this case
\begin{equation}
f_0= -\frac{1}{2\beta}\left\{\ln\left[1+2 e^{\beta(\mu+\delta)}\right] + \ln\left[{1+2 e^{\beta\mu}\, \cosh \left(\beta\eta\right)}\right] \right\},
\end{equation}
and the variational free energy for the SAF phase results:

\begin{widetext}
\begin{eqnarray}
\phi_{SAF}&=&\Phi_{SAF}/N\nonumber \\ 
&=&-\frac{1}{2\beta}\left\{\ln\left[1+2 e^{\beta(\mu+\delta)}\right] + \ln\left[{1+2 e^{\beta\mu}\, \cosh \left(\beta\eta\right)}\right] \right\}
 -J_B\, [1-\rho_A(\delta)]\,m_B^2 +\frac{1}{2}  \eta\, m_B(\eta) + \frac{1}{2} \delta\, \rho_A(\delta)
\label{eq:freeSAF}
\end{eqnarray}
\end{widetext}

The stationarity conditions $\frac{\partial \phi_{SAF}}{\partial \eta}=0$ and
$\frac{\partial \phi_{SAF}}{\partial \delta}=0$ give:
\begin{equation}\label{saf-saddle1}
\delta=-2 J_B m_B^2
\end{equation}
and
\begin{equation}\label{saf-saddle2}
   \eta =4J_B [1-\rho_A(\delta)] m_B.
\end{equation}
Inserting \eqref{saf-saddle1} and \eqref{saf-saddle2} in \eqref{eq:rhoA}, \eqref{eq:mB} and
\eqref{eq:rhoB} we end with three coupled equations for $\rho_A$, $\rho_B$ and $m_B$ to be solved
self consistently. A complete solution can only be possible by solving numerically the set of
equations. Nevertheless, a simple analysis of some limiting cases is useful to check for consistency
of the equations:
\begin{itemize}
\item The global density in the SAF state is defined to be $\rho=(\rho_A+\rho_B)/2$.
\item In the disordered phase, $m_B=0$ implies that $\delta=\eta=0$ and therefore the density is
  uniform:
\begin{equation}
  \rho_A=\rho_B=\frac{2 e^{\beta\mu}}{1+2 e^{\beta\mu}}.
  \label{eq:rhoA=B}
\end{equation}
\item The density saturates in the limit $\mu\to\infty$, $\rho_A=\rho_B=1$, impliying $\eta=m_B=0$, consistently with an antiferromagnetic solution.
\item In the limit $\mu\to -\infty$, $\rho_A=\rho_B=m_B=0$, as expected.
\item There exists a minimum value of the chemical potential $\mu_m$ such that for $\mu<\mu_m$ the only stable phase is the disordered one characterized by $ \langle {\cal O}\rangle =  \langle m_B\rangle=0$.
  For $J_B/J_A=2$, $\mu_m \approx -4.0$.
\end{itemize}
The mean field SAF phase is shown in Figure \ref{fig:meanfield}. From the numerical solution of the
saddle point equations it turns out that the SAF order parameter $m_B$ always changes discontinuously
at the transition line. In the mean field approximation the SAF-disordered transition is discontinous
for any value of the chemical potential $\mu$. This situation changes in the results from Monte Carlo
simulations, as will be shown in the next section. The SAF phase has a dome like shape in the $T-\mu$
or $T-x$ plane. In Figure \ref{fig:meanfield} it can be seen that the AF and SAF phases have a large
coexistence region. Both first order transition lines meet approximately at $(T,\mu)\approx (2.0,0.6)$.
For
$T < 2.0$ the AF and SAF phases are separated by a first order line, which is approximately a straight
line at $\mu =0.6$ in the mean field approximation. Overall, the phase diagram has a similar shape
with
the ones of the iron based high temperature superconductors~\cite{PhysRevLett.104.057006,LI2016925}.

\section{Monte Carlo simulations}

We performed Grand Canonical Monte Carlo simulations using Metropolis algorithm with Hamiltonian \eqref{hamilton} on a square lattice with $N=L\times L$ sites and periodic boundary conditions. For each set of parameters values we let first the system to equilibrate over $2\times 10^4$ Monte Carlo Steps (MCS) and then we average over $M_s$ sample points taken every 100 MCS over a single MC run. $M_s$ run between 1000 and 20000. All the calculation, except those related to hysteresis cycles, were obtained by cooling down at constant steps from high temperature keeping the chemical potential constant and taking the initial configuration at every temperature as the last one of the previous temperature value. We calculated  the average antiferromagnetic staggered magnetization $\langle m_s\rangle$, the orientational order parameter $\langle {\cal O}\rangle$, with ${\cal O}$ given by Eq.(\ref{Ohv}), the associated susceptibilities ($k_B=1$)

\begin{equation}\label{chis}
\chi_s = \frac{N}{T} \left( \langle m_s^2\rangle - \langle m_s\rangle^2\right),
\end{equation}

\begin{equation}\label{chiO}
\chi_O = \frac{N}{T} \left( \langle {\cal O}^2\rangle - \langle {\cal O}\rangle^2\right),
\end{equation}

\noindent and the density

\begin{equation}\label{rhom}
\rho= \frac{1}{N} \sum_{i,j} \langle S_{i,j}^2 \rangle.
\end{equation}

In all cases we used $J_B/J_A=2$.

We obtained the phase diagram in the $(\mu,T)$ plane. We summarize first the main qualitative features of the different regimes observed before presenting a more detailed analysis.

\begin{itemize}
\item There exists a minimum value of the chemical potential $\mu_m$ such that for $\mu<\mu_m$ the only stable phase is the liquid (disordered) characterized by $ \langle {\cal O}\rangle =  \langle m_s\rangle=0$. For $J_B/J_A=2$ and $L=64$, $\mu_m \approx -3.86025$, consistent with the mean field
  result.
\item In the interval between $\mu_m$ and $\mu \approx -1$ a low temperature ordered SAF phase was
  observed, characterized by $ \langle {\cal O}\rangle \neq 0$ and  $\langle m_s\rangle \approx 0$.
  \item According to the value of $\mu$ the phase transition from the SAF to the liquid phase can be of first or second order with the presence of a tricritical point. Namely, for $\mu_m < \mu < \mu_t$ the transition is first order and there is phase coexistence; for $\mu > \mu_t$ the transition is continuous, up to a region around $\mu \approx -1$. The estimated tricritical values for $L=64$ are $\mu_t \approx -2.5$, $x_t \approx 0.5$ and $T_t \approx 1.46$.
  \item For $\mu >0$ there is a stable low temperature AF phase, characterized by $ \langle {\cal O}\rangle \approx 0$ and  $\langle m_s\rangle \neq 0$. The liquid-AF transition is continuous.
  \item Around $\mu =0$ there is a first order phase transition between the ordered phases, with strong hysteresis effects.
\end{itemize}

\begin{figure}[h]
%\centering
\includegraphics[scale=0.52]{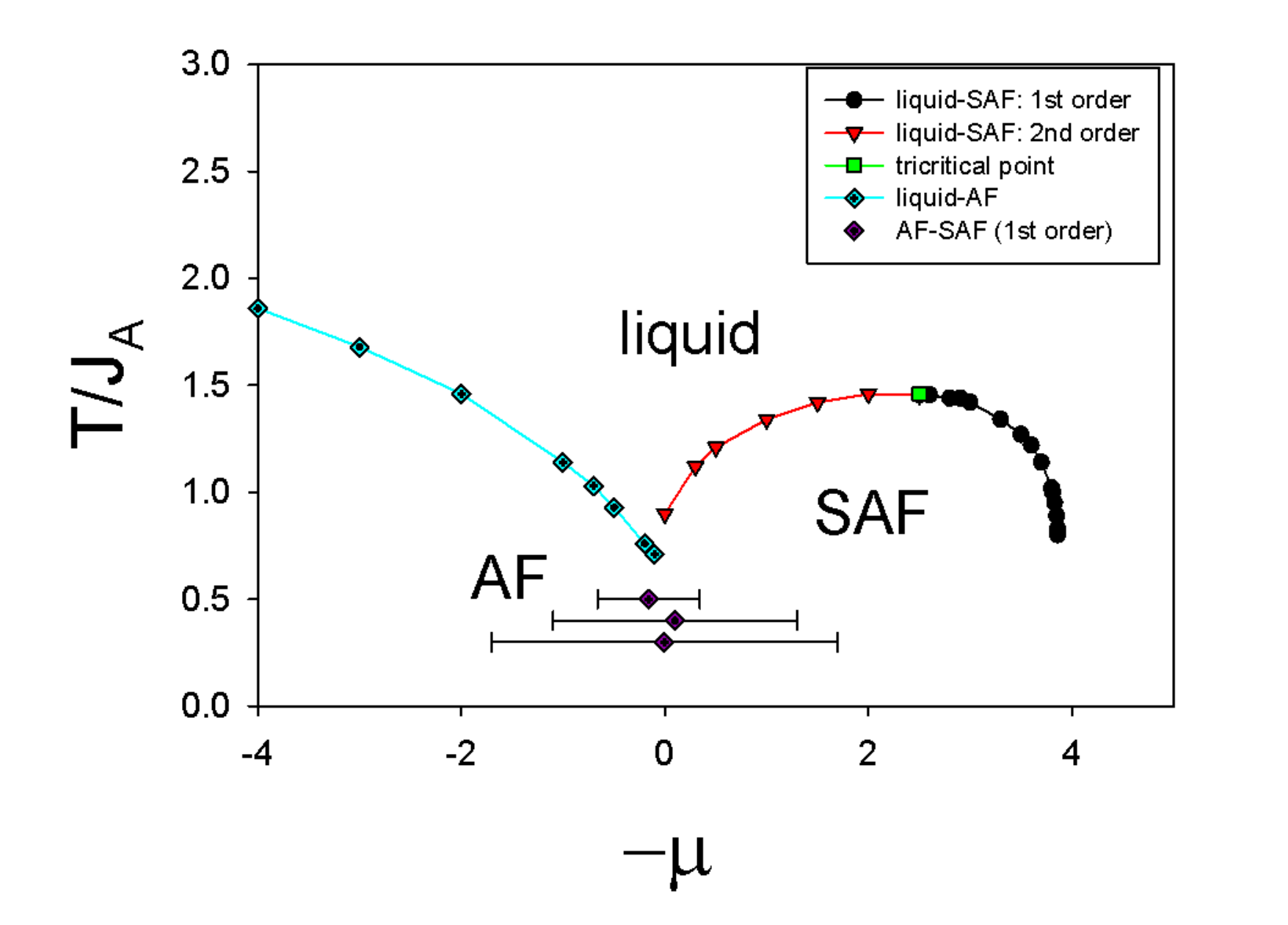}
\caption{(Color online) Temperature vs. chemical potential phase diagram for $J_B/J_A=2$ and $L=64$.}
\label{fig:pd-muT}
\end{figure}

These results are summarized in the phase diagram shown in Fig.\ref{fig:pd-muT}. We next describe the different properties analyzed to obtain that phase diagram.

\begin{figure}[h]
%\centering
\includegraphics[scale=0.5]{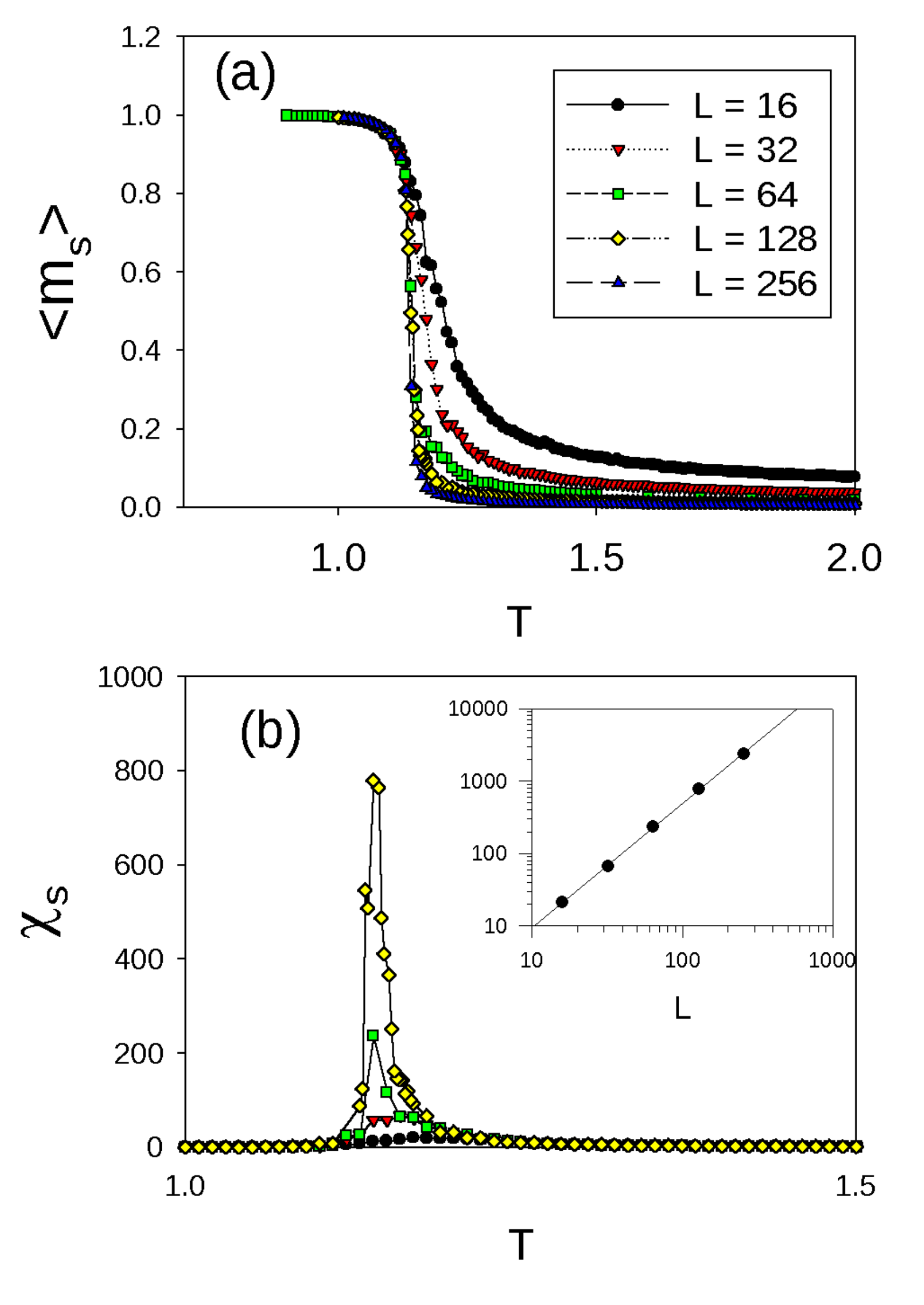}
\caption{(Color online) Second order phase transition behavior for $\mu=1$  and $J_B/J_A=2$. a) AF order parameter (average staggered magnetization) behavior for different values of $L$. b) Associated susceptibility. The inset shows the scaling of the maximum of $\chi_s$ with $L$; a power law fit gives an exponent $\gamma/\nu = 1.72 \pm 0.05$.}
\label{fig:op-2ndorder}
\end{figure}

In Fig.\ref{fig:op-2ndorder} we illustrate the typical  behavior for $\mu \gg 0$. In this region the staggered magnetization becomes different from zero at low enough temperatures, while the orientational order parameter remains almost zero for all temperatures. The finite size scaling behavior of the staggered susceptibility ${\rm max}\; \chi_s \sim L^{\gamma/\nu}$ allows us to estimate the critical exponent $\gamma/\nu = 1.72 \pm 0.05$, in agreement with the exact value for the 2D Ising model $\gamma/\nu = 7/4=1.75$, as expected. At variance with the mean field results, we did not observe evidences of a first order AF-liquid phase transition, nor of a tricritical point. However, large fluctuations close to the region where the transition line joins the SAF-liquid one makes difficult to exclude the existence of a tricritical point in that region. 

\begin{figure}[h]
%\centering
\includegraphics[scale=0.5]{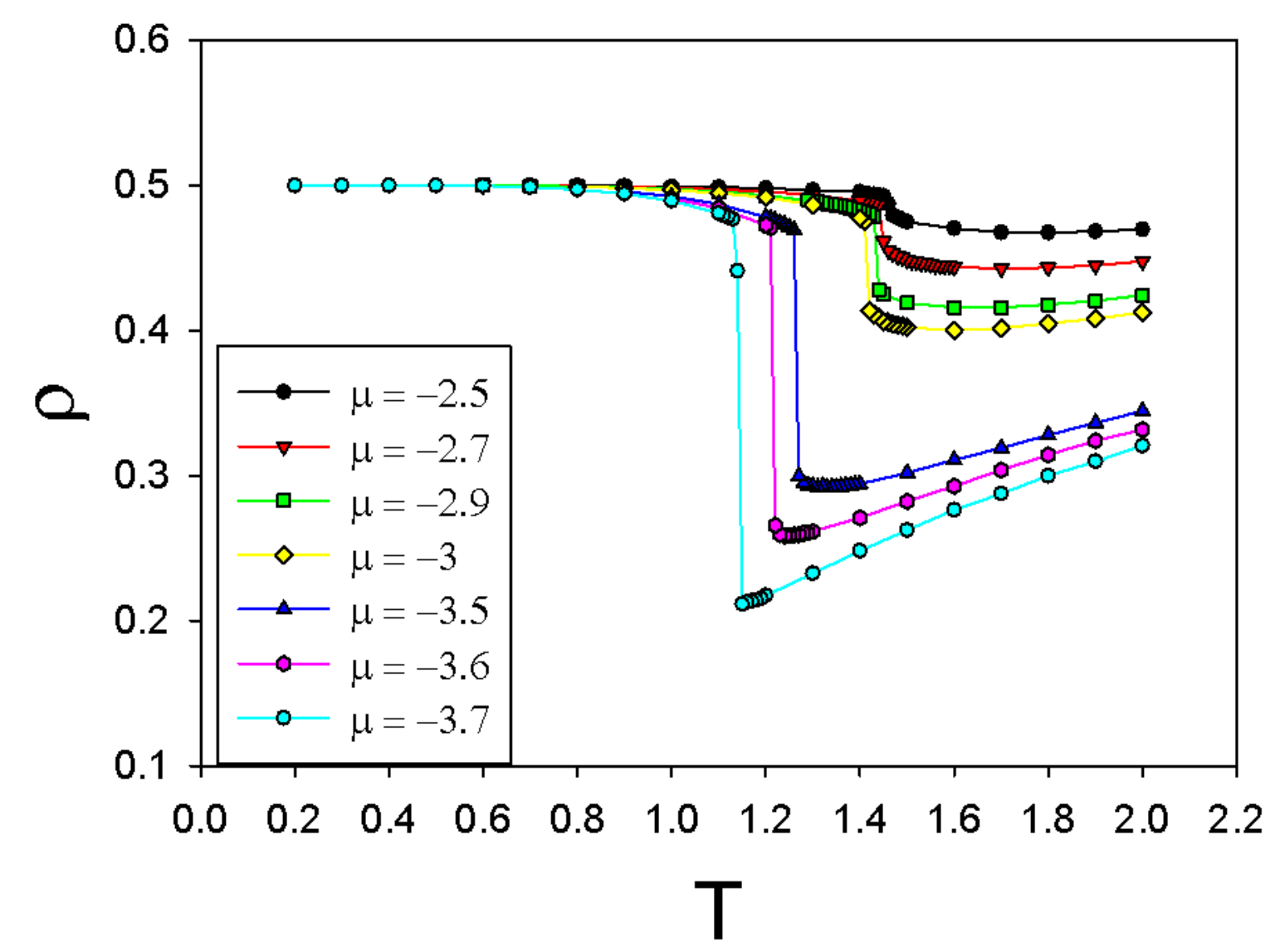}
\caption{(Color online) Density behavior in the first order phase transition between the liquid and SAF phases for $J_B/J_A=2$ and $L=64$.}
\label{fig:dens1storder}
\end{figure}

\begin{figure}[h]
%\centering
\includegraphics[scale=0.5]{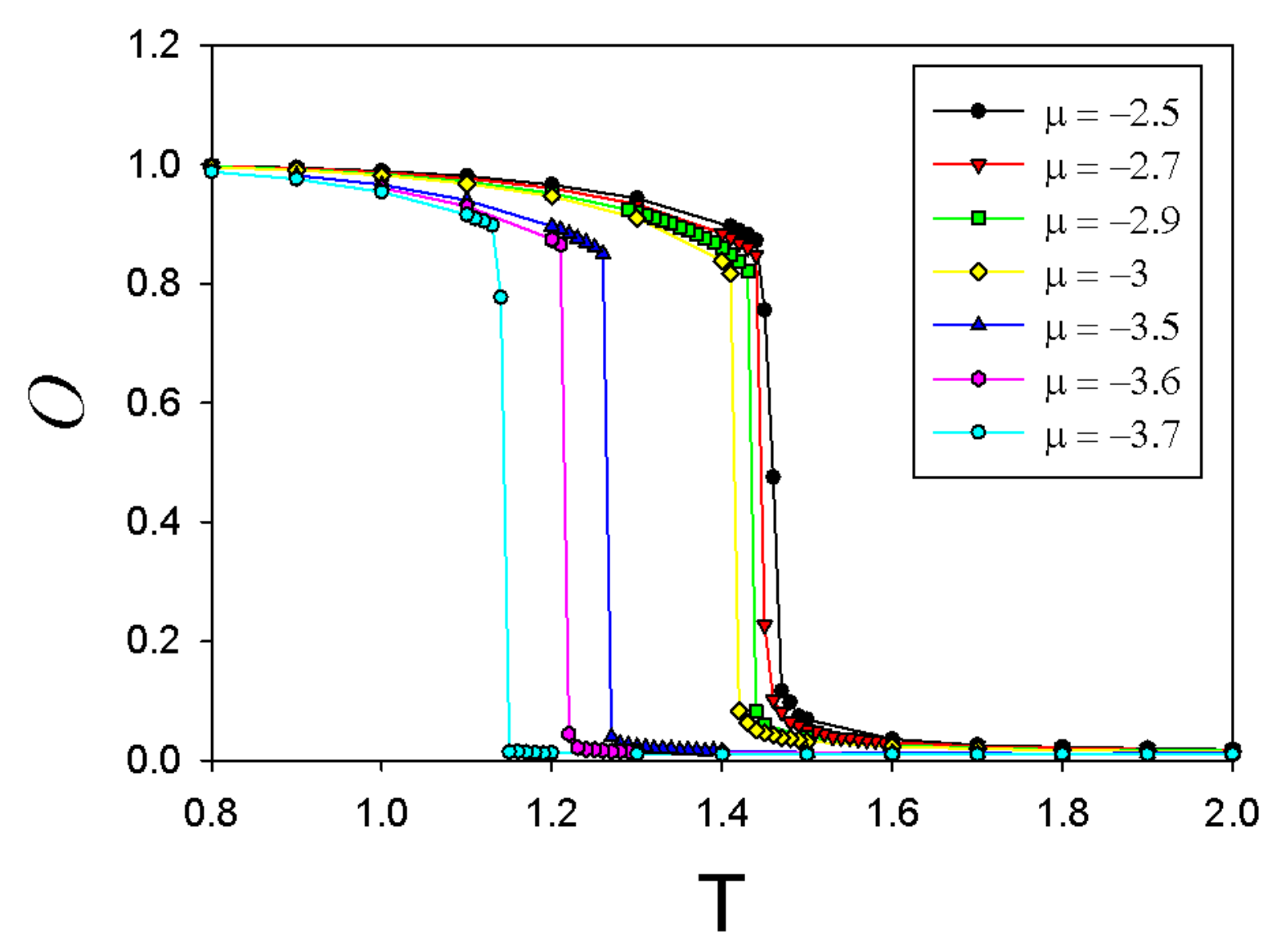}
\caption{(Color online) Orientational order parameter behavior in the first order phase transition between the liquid and SAF phases for $J_B/J_A=2$ and $L=64$.}
\label{fig:op1storder}
\end{figure}

\begin{figure}[h]
%\centering
\includegraphics[scale=0.5]{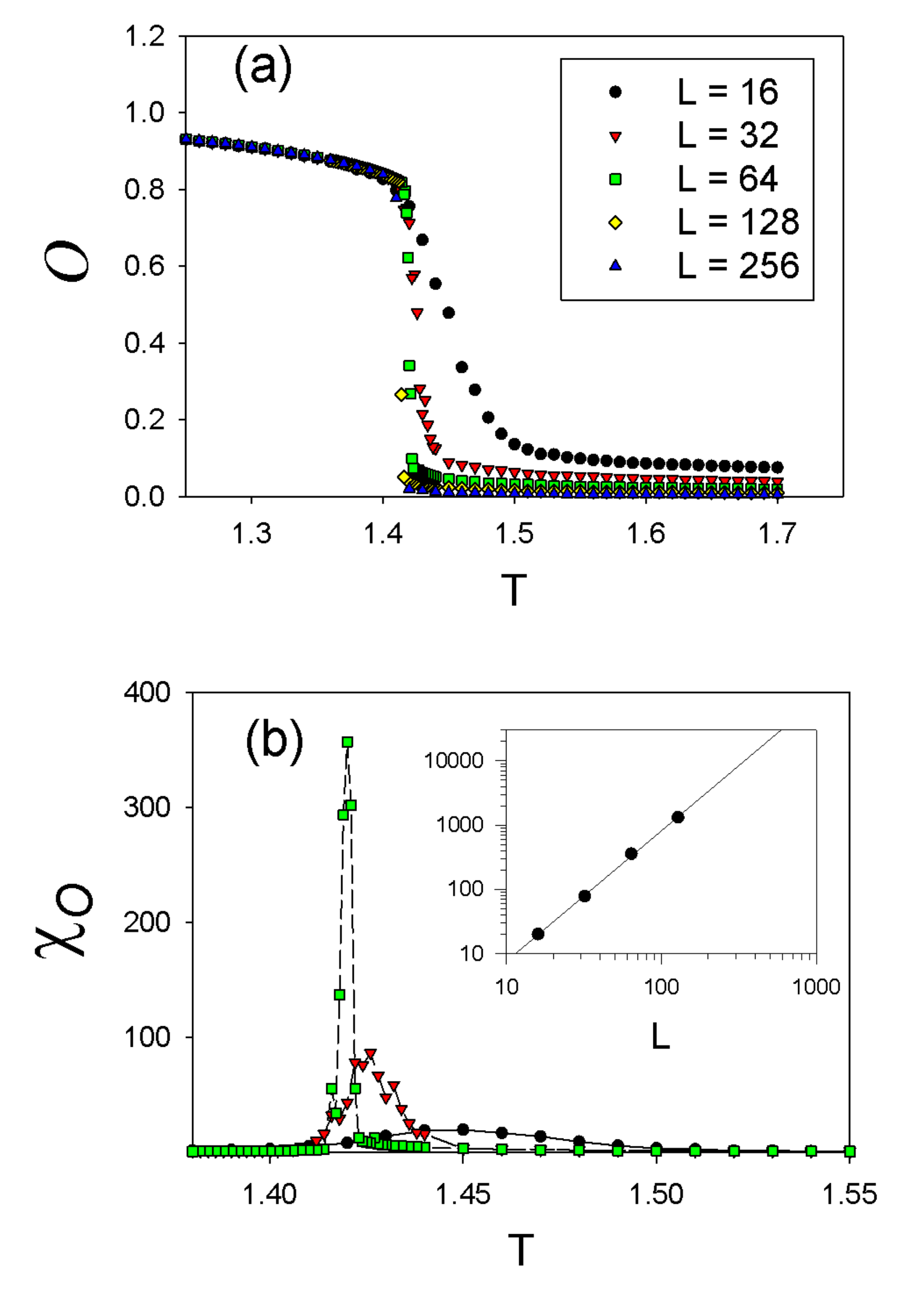}
\caption{(Color online) First order phase transition scaling behavior for $\mu=-3$  for $J_B/J_A=2$. a) Orientational order parameter behavior for different values of $L$. b) Associated susceptibility. The inset shows the scaling of the maximum of $\chi_O$ with $L$; power law fit gives an exponent $\gamma/\nu = 2.03 \pm 0.05$.}
\label{fig:op-mum3}
\end{figure}

Next we analyzed the order parameters behavior for $\mu_m < \mu < \mu_t$. In Figs.\ref{fig:dens1storder} and \ref{fig:op1storder} we show how both the density and the orientational order parameter display a discontinuity that disappears when  $\mu \to \mu_t$. For $\mu > \mu_t$ the density is always continuous.  The  AF order parameter remains almost zero in this region. In Fig.\ref{fig:op-mum3} we show the finite size scaling of the orientational order parameter and the associated susceptibility, whose maximum ${\rm max}\; \chi_O \sim L^{2.03}$ agrees with the expected behavior in a first order transition~\cite{PhysRevB.34.1841} ${\rm max}\; \chi_O \sim L^d$.

\begin{figure}[h]
%\centering
\includegraphics[scale=0.5]{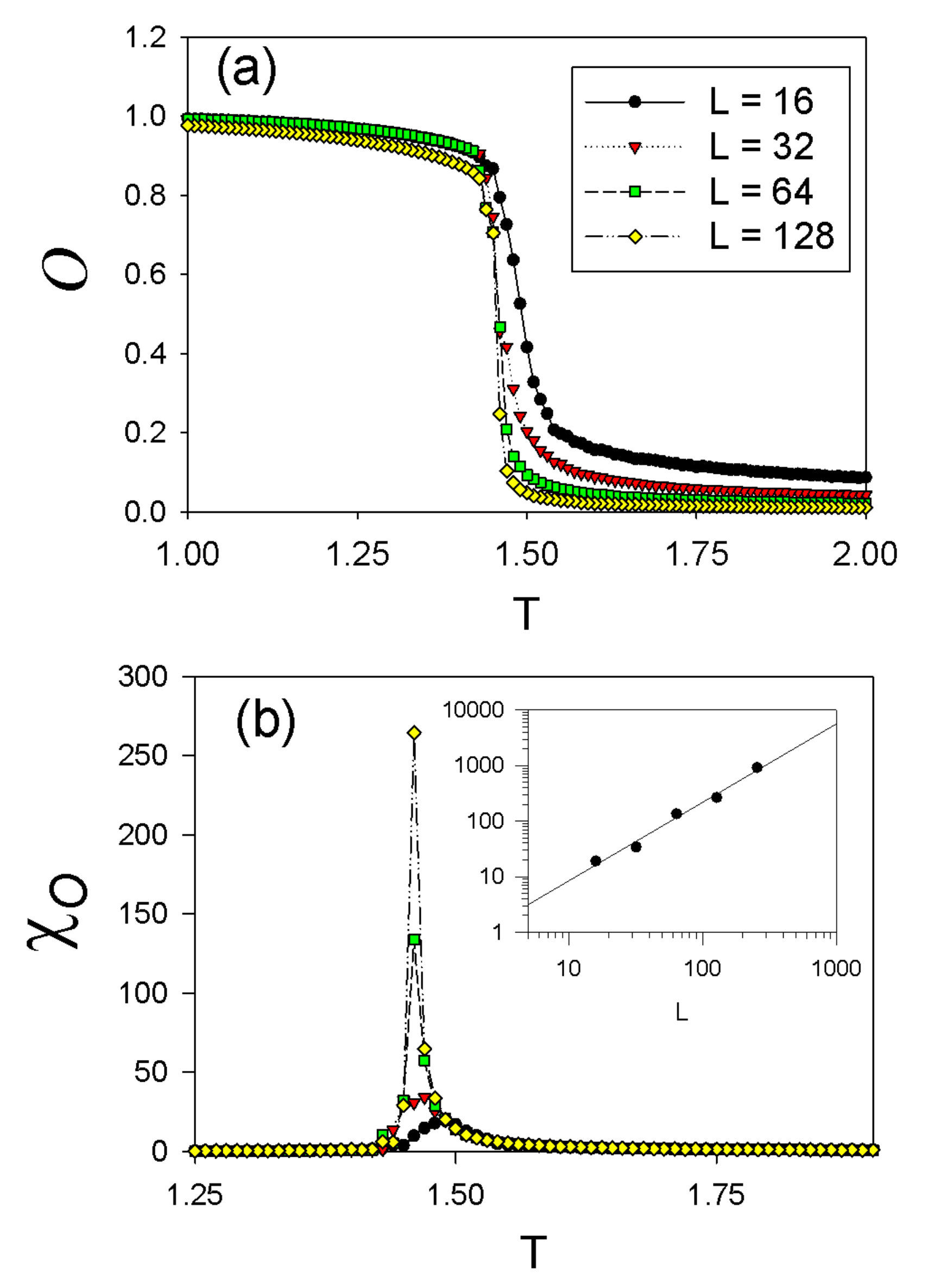}
\caption{(Color online) Second order phase transition scaling behavior for $\mu=-2$  for $J_B/J_A=2$. a) Orientational order parameter behavior for different values of $L$. b) Associated susceptibility. The inset shows the scaling of the maximum of $\chi_O$ with $L$; power law fitting gives an exponent $\gamma/\nu = 1.4 \pm 0.1$.}
\label{fig:op-mum2}
\end{figure}

\begin{figure}[ht!]
%\centering
\includegraphics[scale=0.5]{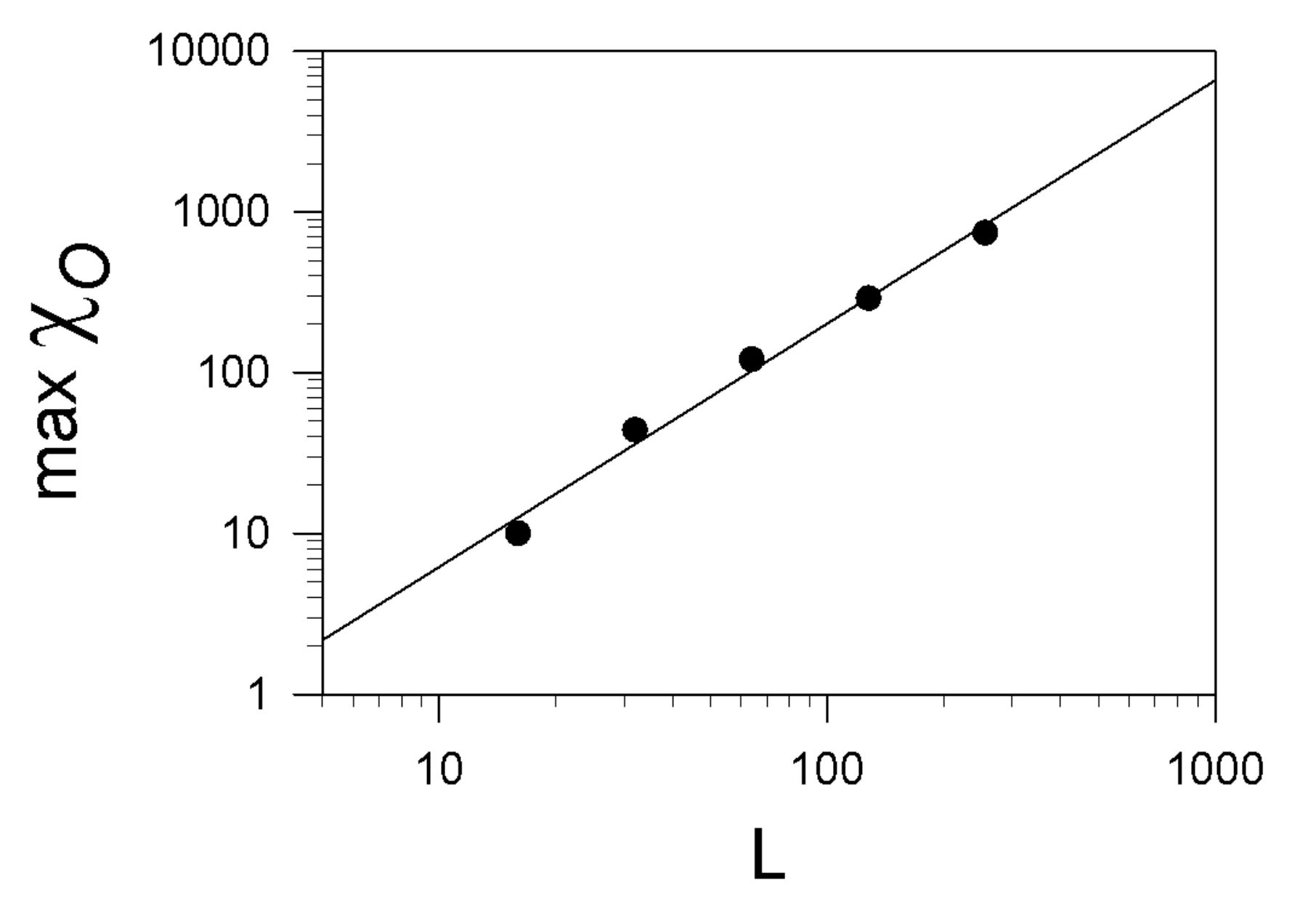}
\caption{Maximum of $\chi_O$ as a function of  $L$ for $\mu=-1$; power law fitting gives an exponent $\gamma/\nu = 1.5 \pm 0.1$.}
\label{fig:chi-mum1}
\end{figure}

In Fig.\ref{fig:op-mum2} we show an example of the finite size behavior of the orientational order parameter in when $\mu >\mu_t$, but not very close to $\mu =0$, where the transition between both ordered phases occurs. The AF order parameter remains almost zero in this region. We see that the maximum of the orientational susceptibility scales with an exponent  $\gamma/\nu \approx 1.4$, well below both from $d$ and from $7/4$, consistently with a continuous phase transition in a universality class different from that of the 2D Ising model. However, this last conclusion should be checked for other values of $\mu$ in the same region, since this value could be influenced by the proximity of the tricritical point. But this is a complicated and tricky analysis, since we have to discriminate  both the finite size and the crossover effects related to the tricritical point. For instance, in Fig.\ref{fig:chi-mum1} we show the finite size scaling behavior of the maximum of $\chi_O$ for $\mu=-1$ and sizes ranging from $L=16$ and $L=256$. If we consider all the points, the power law fitting gives and exponent $\gamma/\nu = 1.5 \pm 0.1$. So, it seems that when we depart from the tricritical point the exponent increases, suggesting an approach to the 2D Ising universality class. However, if we discard the smallest size point we get a better fitting (larger $r^2$) with an exponent $\gamma/\nu = 1.35 \pm 0.07$, which appears to be consistent with the previous value found for $\mu=-2$.

\begin{figure}[ht!] 
%\centering
\includegraphics[scale=0.5]{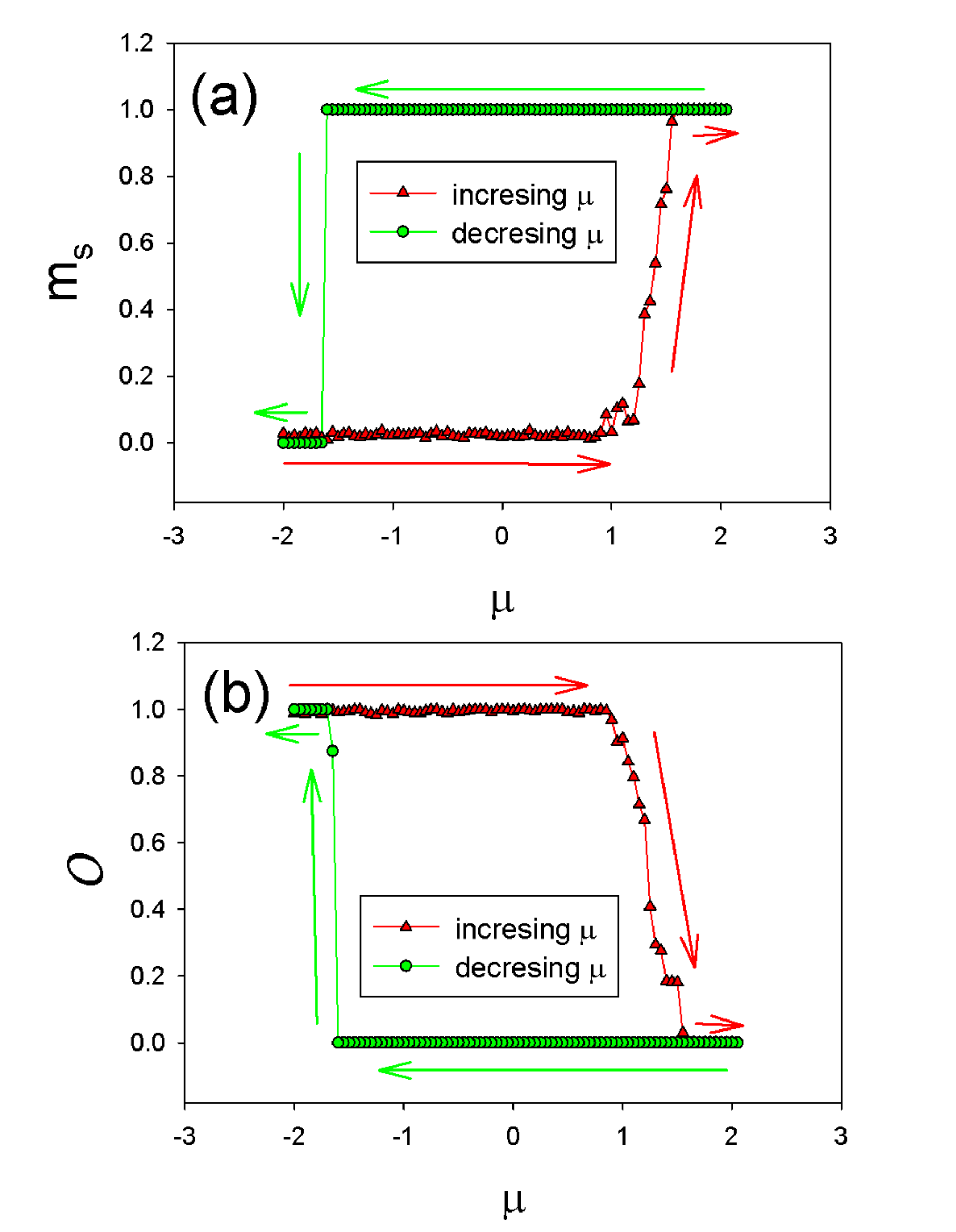}
\caption{(Color online) Order parameters hysteresis cycles for $J_B/J_A=2$, $L=64$ and $T_f=0.4$. (a) Staggered magnetization. (b) SAF order parameter.}
\label{fig:histe}
\end{figure}

Finally, we analyzed the transition from the AF to the SAF phase at low temperature. In general we observed that the behavior in the transition region is very noisy and relaxation times are rather large. Hence, it becomes extremely difficult to determine the nature of the transition, as well as to get a reliable estimation of the transition temperature at fixed chemical potential by means of the standard methods, such as energy and/or order parameters histograms, response functions finite size scaling, etc. Since according to the MF results the transition is a first order one, we can at least check that by looking at hysteresis effects. We then performed chemical potential cycles at constant temperatures in the corresponding region of the phase diagram according to the following protocol. We first let the system to equilibrate $M_e$ MCS at high temperatures in the liquid phase (T=1.5). Then we cool  the system down to a final temperature $T_f$  at constant steps $\Delta T=0.05$ for fixed value of the chemical potential $\mu=\mu_{min}=-2$, letting the system to evolve $M_e$ MCS at every temperature value. The value of $T_f$ was chosen  in such a way that the system starts the cycle well inside the SAF phase. Then we increase the chemical potential at constant $\Delta \mu=0.05$ steps keeping the temperature fixed up to a maximum value $\mu=\mu_{max}=2$ (well inside the AF phase). For every value of $\mu$ we first let the system to evolve $M_e$ MCS and then we calculate the average of both order parameters, taking $10^4$ sampling points every 100 MCS along the same MC run. Once we reach $\mu_{max}$ we repeat all the protocol decreasing $\mu$ down to $\mu_{min}$. Along the whole process we used $M_e = 4\times 10^5$. The results are illustrated in Fig.\ref{fig:histe} for $T_f=0.4$. We observe a strong hysteresis effect, consistently with a first order transition. From these data we got a rough estimation of the transition temperature as an average of the $\mu$ coordinate of  the center of mass of the cycles for every order parameter. The transition points shown in Fig.\ref{fig:pd-muT} were obtained in that way. The associated error bars were estimated as the average half-width of the cycles. Actually, that procedure overestimates the error and those bars rather give an estimation of the location of the spinodal lines.

\section{Discussion}

We introduced an effective lattice-gas model of spin $1/2$ particles doped with holes, to analyze the effect of the competition between an exchange antiferromagnetic interaction and a pairing  interaction mediated by holes. Through this toy model we showed that those two ingredients are enough to reproduce the main features of the global topology of the phase diagram of many high $T_c$ superconducting compounds. Besides some differences in the order of the involved transitions, both mean field and Monte Carlo analysis provide a consistent phase diagram, with a dome in the transition line from the paired induced phase (SAF) and a decreasing AF-liquid transition line as the doping increases (decreasing chemical potential).
In order to turn the model more closely related with the universal characteristics of high $T_c$
superconductors, different modifications and/or generalizations can be envisaged. One possibility is
to include another antiferromagnetic interaction between second neighbours in the square lattice, like in the
$J_1-J_2$ model~\cite{PhysRevLett.64.88,PhysRevB.77.224509}.
The $J_1-J_2$ model has a SAF phase, which is interpreted as spin
or electronic stripes, similar to those observed in the cuprates and pnictides. Furthermore, in
presence of an external field, the model shows a kind of ``electron nematic'' phase, also usually
present in the phase diagram of high $T_c$ compounds~\cite{PhysRevE.91.052123}. The introduction of
holes or doping in the cited models can be an interesting route to explore.

\acknowledgments
DAS acknowledges partial financial support by CNPq, Brazil. This work was partially supported by  CONICET (Argentina) through grant
PIP 11220150100285 and  SeCyT (Universidad Nacional de C\'ordoba, Argentina).

\bibliographystyle{apsrev4-1}
%\bibliography{biblio}
%

\end{document}